\documentclass[aps,pra,showpacs,superscriptaddress,twocolumn]{revtex4-1}
\usepackage{graphicx}  
\usepackage{dcolumn}   
\usepackage{bm}        
\usepackage{amssymb}   
\usepackage{natbib}
\usepackage{amsmath}
\usepackage[colorlinks=true,citecolor=red,linkcolor=blue]{hyperref}
\begin{document}
\title{Theory of Dipole Induced Electromagnetic Transparency}

\author{Raiju Puthumpally-Joseph}
\affiliation{Universit\'e Paris-Sud, Institut des Sciences Mol\'eculaires d'Orsay (CNRS), F-91405 Orsay, France}

\author{Osman Atabek}
\affiliation{Universit\'e Paris-Sud, Institut des Sciences Mol\'eculaires d'Orsay (CNRS), F-91405 Orsay, France}

\author{Maxim Sukharev}
\affiliation{Science and Mathematics Faculty, College of Letters and Sciences, Arizona State University, Mesa, Arizona 85212, USA}

\author{Eric Charron}
\affiliation{Universit\'e Paris-Sud, Institut des Sciences Mol\'eculaires d'Orsay (CNRS), F-91405 Orsay, France}

\date{\today}
\begin{abstract}
A detailed theory describing linear optics of vapors comprised of interacting multi-level quantum emitters is proposed. It is shown both by direct integration of Maxwell-Bloch equations and using a simple analytical model that at large densities narrow transparency windows appear in otherwise completely opaque spectra. The existence of such windows is attributed to overlapping resonances. This effect, first introduced for three-level systems in [R. Puthumpally-Joseph, M. Sukharev, O. Atabek and E. Charron, Phys. Rev. Lett. {\bf 113}, 163603 (2014)], is due to strongly enhanced dipole-dipole interactions at high emitters' densities. The presented theory extends this effect to the case of multilevel systems. The theory is applied to the D1 transitions of interacting $^{85}\text{Rb}$ atoms. It is shown that at high atomic densities, $^{85}\text{Rb}$ atoms can behave as three-level emitters exhibiting all the properties of dipole induced electromagnetic transparency. Applications including slow light and laser pulse shaping are also proposed.
\end{abstract}
\maketitle

\section{Introduction}

 The research of phenomena related to light-matter interaction has been experiencing significant growth for the past decade  \cite{novotny2012principles} in the domain of nano-optics. From the fundamental point of view, one of the most intriguing question lies in the dynamics and in the optical properties of many-body systems. For instance, new collective modes were predicted and measured in materials composed of interacting quantum emitters (atoms, molecules, etc.)  \cite{PhysRevLett.109.073002}. Experiments in that regard are noticeably advanced compared to theory, as it has long been realized that treating multi-body systems is quite challenging. Even systems comprised of just a few interacting quantum dipoles are difficult to investigate  \cite{PhysRevA.62.013413, PhysRevLett.112.113603}. Semi-classical descriptions of light-matter interaction, in which a classical electric field interacts with a quantum system,  enormously simplify modeling leading to results that both support experiments and predict new phenomena \cite{doi:10.1021/nn4054528,PhysRevLett.109.073002,PhysRevLett.113.163603}.
 
 Collective effects \cite{PhysRev.93.99} arise from the fact that  the electric field experienced by an arbitrary quantum emitter is the sum of both the applied field and the radiation from all other atoms in the system under consideration. When it is strong enough, the radiation from the dipoles set them coupled. This is the source of cooperativity.  One of the important fundamental cooperative effects, the collective Lamb shift which is due to the exchange of virtual photons between the particles in a dense sample of atoms was recently observed \cite{PhysRevLett.108.173601}. Dependent of the cooperativeness in homogeneous and inhomogeneous systems were studied both experimentally \cite{PhysRevLett.104.183602,PhysRevLett.113.133602} and theoretically \cite{PhysRevLett.112.113603, PhysRevA.61.063814, PhysRevA.61.063815}.   At present, research on those cooperative effects concerns on both cold \cite{PhysRevLett.112.040501,PhysRevLett.101.103602,nat.390.671,PhysRevLett.112.023905,PhysRevLett.113.223601} and thermal \cite{PhysRevLett.108.173601,arXiv:1308.0129,PhysRevLett.109.233001} samples of atoms since they offer the observation of different interesting quantum phenomena. Because of the ability of light to carry effectively  quantum information, understanding the optical response and the light scattering in dense samples is also interesting for many applications such as slow and stopped light \cite{nature09933,PhysRevLett.108.173603}, quantum information \cite{nature07127,PhysRevLett.87.037901}, trapped light \cite{Trapped_rainbow} and optical memories \cite{PhysRevLett.103.233602}.
 
In the linear regime, the propagation of light through any medium can be investigated by extracting its electric susceptibility \cite{jackson1962classical, j.phys.b.41.155004}. It was  shown recently \cite{PhysRevLett.113.163603} that the electric susceptibility of systems comprised of three level quantum emitters at high densities exhibit a Fano-type profile that results in a new phenomenon of induced transparency called \emph{Dipole Induced Electromagnetic Transparency} (DIET). DIET is due to the collective response of the system towards the external electromagnetic (EM) field via dipole-dipole interactions. Following this work, in this paper, we present a detailed theory of DIET and extend the model to multilevel systems. We also apply our theory to a realistic system, namely the D1 line of $^{85}\text{Rb}$.

The paper is organized as follows. The next section discusses in detail our theoretical description for $N$-level systems and the analytic model used to analyze the results. Section \ref{two-levels} provides results and a comprehensive discussion of how linear optical properties of a system comprised of interacting two-level emitters depends on various material parameters such as the particle density. In Section \ref{many-levels}, we extend our discussion to systems having more than one excited states. First we consider interacting three-level emitters. The results provided demonstrate dipole induced electromagnetic transparency (DIET). This phenomenon is explained using an analytical model and is extended to the multilevel D1 transitions of $^{85}\text{Rb}$.  Section \ref{applications} discusses possible applications of our results and finally, we conclude our work in Section \ref{conclusion}.

\section{Theoretical models}
\label{model}

Our goal is to describe the linear dynamics of quantum emitters in a self-consistent manner taking into account collective effects. The most comprehensive and thus complete description would be to implement a fully quantum model based on the Jaynes-Cummings Hamiltonian \cite{doi:10.1080/09500349314551321}. On the other hand, the quasi-classical theory of light-matter interaction recently applied to hybrid materials has proven to lead to both qualitative and quantitative agreement with experiments  \cite{torma2014strong}. Keeping also in mind that the majority of experiments in nano-optics, such as white light spectroscopy, for instance, are performed under conditions corresponding to high occupation numbers of photons in a given EM mode, the quasi-classical description is well justified.

The idea is to separate the description of EM waves from the dynamics of quantum emitters considering the spatio-temporal evolution of EM radiation using classical Maxwell's equations while applying the full machinery of quantum mechanics to describe the response of quantum emitters to EM excitation. This results in a system of coupled Maxwell-Bloch equations  \cite{PhysRevA.84.043802}. If one wants to observe collective effects, it is imperative to solve the corresponding equations of motion self-consistently without any decoupling  \citep{jcp1.4774056}. 
 
\subsection{Theoretical Model and Numerical Simulations}

  The dynamics of the EM field components, $\vec{E}$ and $\vec{H}$, is simulated using Maxwell's equations
  \begin{subequations}
   \label{Maxwell}
    \begin{align}
     & \mu_0 \frac{\partial \vec{H}}{\partial t}  =  -\nabla \times \vec{E} \\
     & \varepsilon_0\frac{\partial \vec{E}}{\partial t}  =  \nabla \times \vec{H} - \frac{\partial\vec{P}(\vec{r},t)}{\partial t}, 
    \end{align}
  \end{subequations}
 where $\varepsilon_0$ and $\mu_0$ are the permittivity and the permeability of free space, respectively, and $\vec{P}(\vec{r},t)$ is the 
macroscopic polarization of the system at position $\vec{r}$ and time $t$. The dynamics of the latter is considered both quantum mechanically and classically. In both cases we assume that the emitters are continuously distributed in space and we neglect static emitter-emitter interactions. Under such conditions one can express the macroscopic polarization as
  \begin{equation}
   \label{mean_field_approximation}
    \vec{P}(\vec{r},t) = n_0 \langle\vec{\mu}\rangle,
  \end{equation}
where $n_0$ is the number density of emitters. The average dipole moment, $\langle\vec{\mu}\rangle$, is either determined directly from classical equations of motion or quantum mechanically by evaluating
\begin{equation}
 \label{polarization_matrix_trace}
 \langle\vec{\mu}\rangle = \text{Tr}\big{[}\hat\rho (\vec{r},t)\vec{\mu}\big{]},
\end{equation}
where $\hat\rho (\vec{r},t)$ denotes the density matrix of the system. Each quantum emitter has $(N-1)$ excited states represented by $|\jmath\rangle$, with $\jmath\geq 1$ and 
  $N \geq 2$, coupled to the ground state $|0\rangle$ via the time dependent EM interaction $V_{\text{int}}(\vec{r},t)$. The density matrix $\hat\rho \left(\vec{r},t\right)$ satisfies the following Liouville-von Neumann equation 
  \begin{equation}
   \label{Opt_Bloch_1}
    i\hbar\frac{\partial\hat\rho}{\partial t} = [\hat {\cal {{\cal {H}}}}, \hat\rho]-i\hbar\hat\Gamma\hat\rho,
  \end{equation}
where $\hat {{\cal {H}}}= \hat {\cal {H}}_0 + V_{\text{int}}(\vec{r},t)$ is the total Hamiltonian and $\hat\Gamma$ is the superoperator describing relaxation and dephasing processes taken in the Lindblad form under Markov approximation  \cite{petruccione2002theory}. Off-diagonal elements of $\hat\Gamma$ include the pure dephasing rate $\gamma^*$ and the diagonal elements include the radiationless decay rate $\Gamma$ of the excited states.

The field free Hamiltonian $\hat {\cal {H}}_0$ is given by

\begin{equation}
 \label{field_free_Hamiltonian}
  \hat {\cal {H}}_0 =  \sum_{\jmath=0}^{N-1} \hbar \omega_{\jmath }| \jmath \rangle \langle \jmath |.
\end{equation}

We define the Bohr frequencies as $\omega_{0\jmath}=\omega_{\jmath }-\omega_{0}$ and the interaction of a single emitter with the EM fields  is written as

  \begin{equation}
   \label{interaction}
    \hat V_{\text{int}}\left(\vec{r},t\right) = \sum_{\jmath = 1}^{N-1} \hbar\Omega_{\jmath}\left(\vec{r},t\right)\left(|0\rangle \langle \jmath | + | \jmath \rangle \langle 0 |\right),
  \end{equation}
 where $\{\Omega_{\jmath}(\vec{r},t)\}$ are the instantaneous Rabi frequencies associated with the interaction between the quantum system and the local EM fields. In the present work we assume that the excited states are not directly coupled to each other but only to the ground state $|0\rangle$. The corresponding system of coupled differential equations reads  \cite{allen2012optical}
 
  \begin{subequations}
   \label{Opt_Bloch_2}
    \begin{align}
     \dot\rho_{00} & = \sum_{\jmath\geq1}i\Omega_{\jmath}(\vec{r},t)(\rho_{0\jmath}-\rho_{\jmath 0}) + \Gamma\rho_{\jmath\jmath}, \\
     \dot\rho_{\jmath\jmath} & = i\Omega_{\jmath}(\vec{r},t)(\rho_{\jmath 0}-\rho_{0\jmath}) - \Gamma\rho_{\jmath\jmath}, \\
     \dot\rho_{0\jmath} & = i\Omega_{\jmath}(\vec{r},t)(\rho_{00}-\rho_{\jmath\jmath}) + \Big[ i\omega_{0\jmath}-\gamma \Big]\rho_{0\jmath},
   \end{align}
  \end{subequations}
where

\begin{equation}
\label{decay_rate}
 \gamma = \frac{2\gamma^*+\Gamma}{2}.
\end{equation}

 Eqs. (\ref{Maxwell}) and (\ref{Opt_Bloch_1}), coupled through the evolution of the quantum polarization (\ref{mean_field_approximation}), form the basis of the model. As this system of equations is propagated in space and time on a grid, one should note that each grid point is effectively a point-wise dipole with the amplitude of an individual emitter times the number density. This essentially means that dipole-dipole interactions within a single grid point are neglected. This approximation although valid at low densities has to be corrected for high densities. Treating the dipoles exactly to include those contributions is extremely difficult and it is almost impossible to solve the system of equations  by using current computational facilities \cite{PhysRevLett.112.113603}. An alternative way to include dipole-dipole interactions at a single grid point level is to introduce the well-known mean field Lorentz-Lorenz correction term to the local field  \cite{jackson1962classical}, with   
  \begin{equation}
   \label{local_field}
    \vec{E}_{\text{local}} = \vec{E} + \frac{\vec{P}}{3\varepsilon_0}.
  \end{equation}
 
This local electric field enters the Liouville-von Neumann equation (\ref{Opt_Bloch_2}) through the Rabi frequencies $\Omega_{\jmath} (\vec{r},t) = \mu_{0\jmath} E_{\text{local}}(\vec{r},t)/\hbar$, where $\mu_{0\jmath}$ denotes the transition dipole moment between states $|0\rangle$ and $|\jmath \rangle$. 

\begin{figure}[h]
     \includegraphics[width=8.5cm]{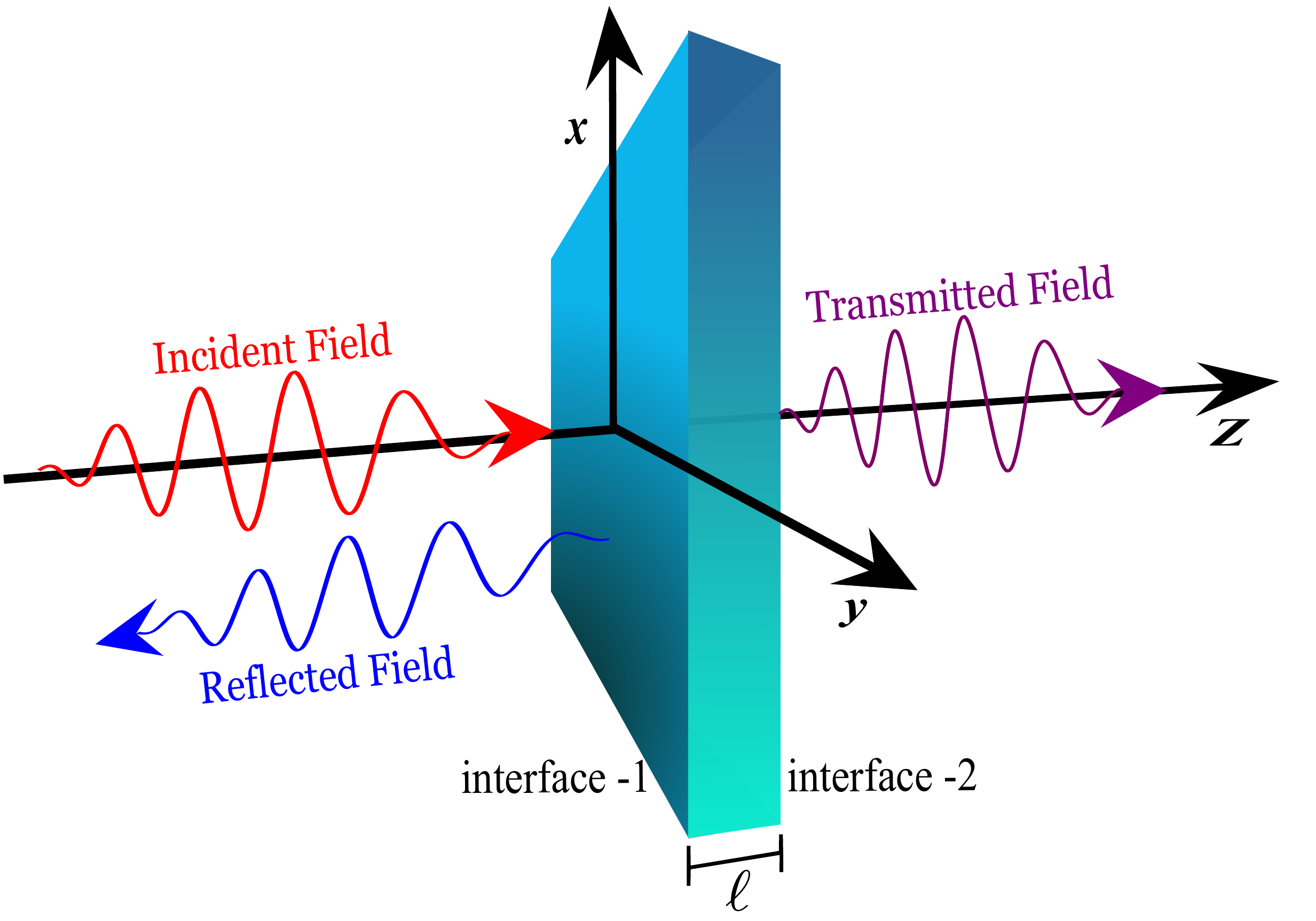}
      \caption{(Color online) The schematic setup of simulations: a slab of thickness $\ell$ composed of interacting quantum emitters is exposed to linearly polarized EM fields acting normal to the interface. }
   \label{system}
   \end{figure}

Using the proposed model we simulate the linear optical response of a vapor comprised of interacting quantum emitters as schematically depicted in Fig. \ref{system}. To simulate a first order elastic scattering/absorption we implement the short pulse method \cite{PhysRevA.84.043802}. A weak ultra-short incident pulse polarized along $\hat{x}$ is launched at normal incidence ($\hat{z}$ direction) on a slab of quantum emitters of finite thickness. 

For such a geometry, Maxwell's equations (\ref{Maxwell}) reduce to
   \begin{subequations}
    \label{reduced_maxwell}
     \begin{align}
      & \mu_0\frac{\partial H_y}{\partial t} = -\frac{\partial E_x}{\partial z}, \\
      & \varepsilon_0 \frac{\partial E_x}{\partial t} =-\frac{\partial H_y}{\partial z}-\frac{\partial P_x}{\partial t}.
     \end{align}
   \end{subequations}
 
Maxwell's equations are numerically integrated using a finite-difference time-domain (FDTD) technique  \cite{taflove2000computational} with a spatial grid step, $\delta z = 1$ nm and temporal grid step, $\delta t = 1.67$ as. Convolutional perfectly matched layers (CPML)  \cite{CPML_paper} absorbing boundaries of thickness $19$ nm are used to avoid non-physical reflections from the boundaries of the simulation region. Concurrently, Liouville-von Newmann's equations (\ref{Opt_Bloch_2}) are propagated in time using the fourth order Runge-Kutta method, assuming that all emitters are initially in their ground state $|0\rangle$. Light excitation is provided by a gaussian pulse of FWHM $\delta \omega = 2\pi/\tau$, where $\tau$ is the pulse duration.
   
 To evaluate transmission, reflection and extinction, we calculate the EM energy flux  $ \tilde{S}(\omega)$ on the input and the output sides of the slab using
 
  \begin{equation}
   \label{Poynting2}
    \tilde{S}(\omega) = \frac{\tilde{{E}}_{x} \tilde{{H}}_{y}}{|\tilde{{E}}_{x,\text{inc}} \tilde{{H}}_{y,\text{inc}}|},
  \end{equation}
where $\tilde{E}_{x}$, $\tilde{{H}}_{y}$, $ \tilde{E}_{x,\text{inc}}$ and $\tilde{H}_{y,\text{inc}}$ are the Fourier components of the total and incident EM fields. Eq. (\ref{Poynting2}) gives the reflection spectrum $R(\omega)$ if $\tilde{E}_{x}$ and $\tilde{{H}}_{y}$ are the reflected EM fields and the transmission spectrum $T(\omega)$ if $\tilde{E}_{x}$ and $\tilde{{H}}_{y}$ are the transmitted EM fields. Both spectra carry the information about the dynamics of the system. 

And finally the extinction $A(\omega)$ is estimated as 
 
  \begin{equation}
   \label{absorption}
    A(\omega) = 1-\left[T(\omega)+R(\omega)\right].
  \end{equation} 
  
The layer of quantum emitters is placed at a distance of $1.25$ $\mu$m from the source. The reflected field is measured at a point between the source and the layer which is at $10$ nm away from the source and the transmitted field is measured on the output side, $2.5$ $\mu$m far from the source. We verified that the results are independent of those distances as long as the pulse acts normal to the layer.

\subsection{Semi-classical Approximation}
 \label{Semi-classical Approximation}    
 In the linear regime, $\rho_{00} \gg \{\rho_{{\jmath}{\jmath}}\}$, for ${\jmath}\geq1$, and the homogeneous system of $N$-level quantum emitters can be described by a set of $(N-1)$ coupled harmonic dipole oscillators. The macroscopic polarization associated with the transition between $|0\rangle$ and $|\jmath \rangle$, $P_{\jmath}$, follows a classical equation of motion
 \begin{equation}
  \label{damped oscillator_corrected1}
   \ddot {P}_{\jmath}(t)+\gamma_{cl}\dot{P}_{\jmath}(t)+\omega_{0\jmath}^2 P_{\jmath}(t) =  \bigg{(}\dfrac{n_0q_{\jmath}^2}{m_\jmath}\bigg{)} E_{\text{local}}(t),
 \end{equation}
where $\gamma_{cl} = 2\gamma$ and $m_{\jmath}$ is the effective mass of the oscillating charge $q_{\jmath}$ driven by the  electric field $E_{\text{local}}(t)$.  Taking into account Eq. (\ref{local_field}) and subsequently applying a Fourier transform, we obtain

 \begin{equation}
  \label{damped oscillator_corrected2}
   W_{\jmath}(\omega) \tilde{P_{\jmath}}(\omega) = \varepsilon_0 \omega_{{\cal P} \jmath}^2 \tilde{E_x}(\omega)+ \frac{\omega_{{\cal P} \jmath}^2}{3}\sum_{k}\tilde{P_{k}}(\omega),
 \end{equation}
 where

 \begin{equation}
  \label{frequency_part}
  W_{\jmath}(\omega) =  \omega_{0\jmath}^2 -\omega^2 +i\gamma_{cl}\omega
 \end{equation}
 and where $\omega_{{\cal P} \jmath}$ is the plasma frequency representing the collective oscillations of the dipoles in the system 

 \begin{equation}
  \label{lorentz_shift1}
   \omega^2_{{\cal P} \jmath} = \frac{n_0q_{\jmath}^2}{\varepsilon_0m_{\jmath}}.
 \end{equation}
 
 With the assumption that the maximum amplitude of dipole oscillations in the absence of a driving field is given by the quantum harmonic oscillation length, we have

  \begin{equation}
  \label{lorentz_shift2}
   \omega^2_{{\cal P} \jmath} = \frac{2\omega_{0\jmath}n_0\mu_{\jmath}^2}{3\hbar\varepsilon_0}.
 \end{equation}

 In weak electric fields where $\tilde{P_{\jmath}}(\omega)  = \chi_{\jmath}\varepsilon_0 \tilde{E_x}(\omega)$, Eq.   (\ref{damped oscillator_corrected2}) gives the susceptibility $\chi_{\jmath}(\omega)$ associated with the $\jmath^{\text{th}}$ dipole. Summing over all dipole oscillators in the material, we obtain the total susceptibility of the system
 
\begin{subequations}
  \label{chi_in_coupled_N_level_systems}
   \begin{align}
   X(\omega) & = \sum_{\jmath}\chi_{\jmath}(\omega) = \frac{{\sum_{\jmath}{\omega^2_{{\cal P} \jmath}/W_{\jmath}(\omega)}}}{{1-\frac{1}{3}\sum_{\jmath}{\omega^2_{{\cal P} \jmath}/W_{\jmath}(\omega)}}}.
    \end{align}
 \end{subequations}
 
  \section{Interacting two-level emitters}
  \label{two-levels}

First we consider a slab of interacting simple two-level emitters characterized by the transition energy $\hbar\omega _{01}= 2$ eV, the transition dipole $\mu_1 = 2$ D, and the total decoherence rate $\gamma = 10.5$ THz. We calculate the response  of the system to an extremely short pulse of width $\tau = 0.18$ fs using coupled Maxwell-Liouville-von Neumann equations in different collective interaction regimes  \cite{j.phys.B:44:195006}. The results are analyzed using the semi-classical approximation described in Section \ref{Semi-classical Approximation}.

\subsection{Semi-classical approximation and the Spectra}
 The electric susceptibility of interacting two-level emitters in the semi-classical model can be written as 
 
  \begin{equation}
  \label{chi1}
   X(\omega) = \frac{\omega^2_{{\cal P} 1}}{\tilde\omega_{01}^2 -\omega^2 +i\gamma_{cl}\omega  },
 \end{equation}
 where $\tilde{\omega}_{01}=\sqrt{\omega_{01}^2-\frac{1}{3}\omega_{{\cal P} 1}^2}$ is the shifted resonant frequency of the system. The shift in transition frequency, $\Delta_{1}$, which is due to the Lorentz-Lorenz correction included in the formulation of the local electric field via Eq.  (\ref{local_field}) can be estimated from the semi-classical model. The first order term gives the well-known Lorentz-Lorenz shift (LL shift) \cite{jackson1962classical,lorentz1916theory,PhysRevA.61.063815}. 
 
 \begin{equation}
  \label{shift_delta}
   \Delta_{1}= n_0\mu_{1}^2/9\hbar\varepsilon_0,
 \end{equation}
 
The case of strong dipole-dipole interactions is characterized by $\Delta_1 \gg \gamma$, while $\Delta_1 \ll \gamma$ corresponds to the weak interaction regime.
 
\begin{figure}[h]
   \includegraphics[width=8.5cm]{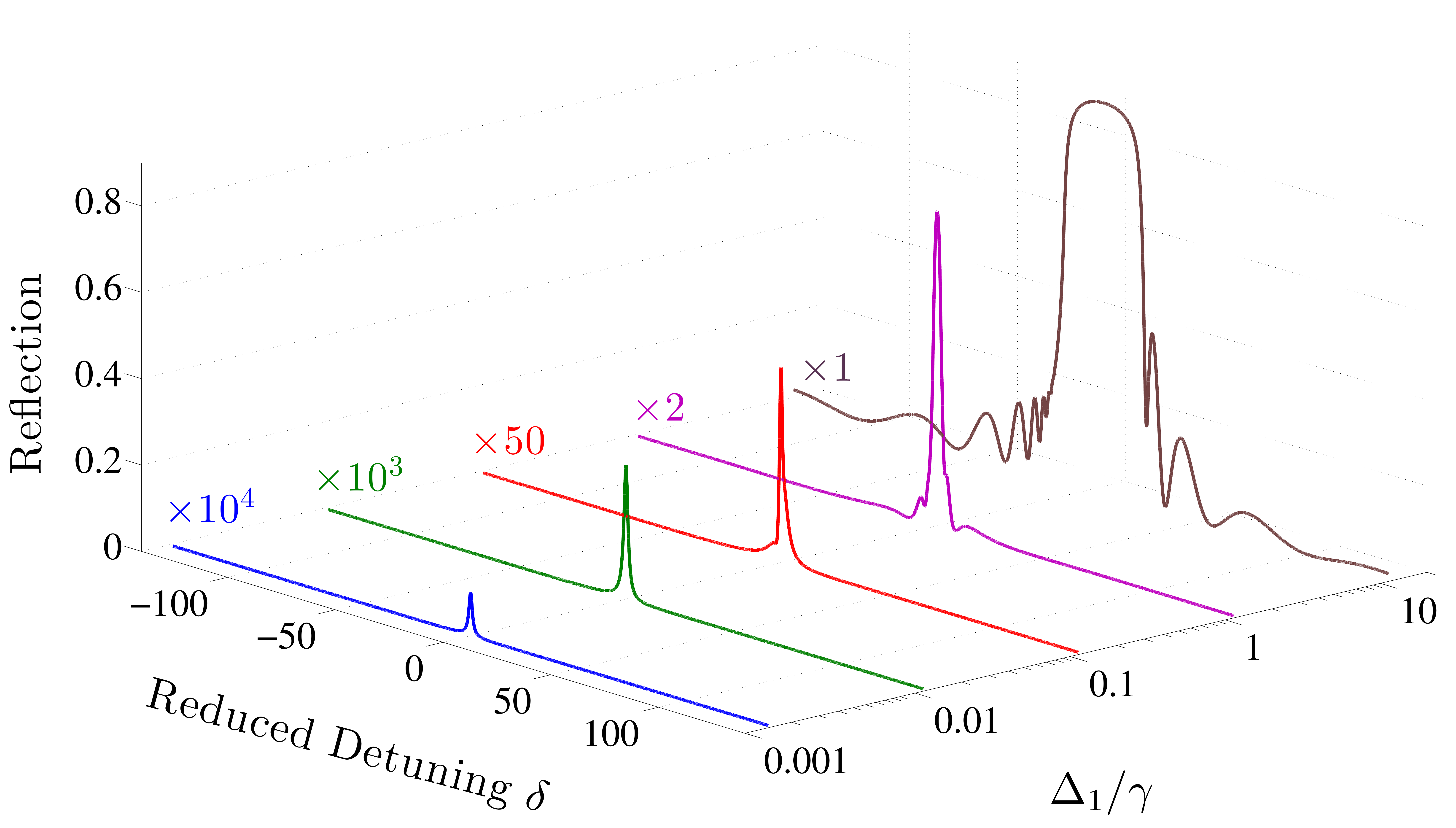}
    \caption{(Color online) Reflection spectra as a function of the  reduced detuning $\delta = (\omega-\omega_{01})/\gamma $ for a slab of interacting two-level  emitters in different interaction regimes. The thickness is $\ell= 600$ nm and the transition dipole $\mu_{1}= 2$ D. The radiationless decay rate and the pure dephasing rate are $\Gamma = 0.5$ THz and $\gamma^* = 10$ THz , respectively. Reflection for small values of $\Delta_{1}/\gamma$ is weak and hence amplified (for better comparison) by the constant factors shown in the Figure. }
   \label{ref_two_level}
  \end{figure}

 Typical reflection spectra are shown in Fig.  \ref{ref_two_level} as a function of the reduced detuning, $\delta = (\omega-\omega_{01})/\gamma$, for different values of $\Delta_{1}/\gamma$ for a slab of thickness $\ell=600$ nm. In the weak interaction regime ($\Delta_{1}/\gamma = 10^{-3}$), the system reflects a very small amount of the total energy (of the order of $10^{-5}$) around the transition frequency. Such a response is similar to that of a single emitter since it is characterized by a typical Lorentzian profile. As the value of $\Delta_{1}/\gamma$ increases, and especially for $\Delta_1 > \gamma$, the interaction between the dipoles becomes dominant and the system starts to respond to the applied field collectively. For $\Delta_{1}/\gamma \geq 10$, the reflection spectrum broadens resulting in a window of frequencies within which more than $80\%$ of  incident energy is reflected. This reflection window was explained in $2000$ by R. J. Glauber and S. Prasad in Ref. \cite{PhysRevA.61.063814, PhysRevA.61.063815} by considering the exponentially decaying modes of coherent excitation of the medium that depend strongly on the frequency.
 
   For a slab of thickness $\ell$ larger than the dipole wavelength $\lambda_{01} = 2\pi c/\omega_{01}$, the reflection at the vacuum/slab interface (shown as interface $1$ in Fig. \ref{system}) dominates and the contribution from the possible multiple reflections inside the slab can be neglected. For such a slab, the width of the reflection window can be estimated from the reflection probability at this interface.
   
 \begin{equation}
  \label{Reflectivity}
   {\cal{R}}(\omega)=  \bigg{|}\frac{1-{\cal N}(\omega)}{1+{\cal N}(\omega)}\bigg{|}^2 ,
 \end{equation}
where ${\cal N}(\omega) = \text{Re}[\sqrt{1+X(\omega)}]$ is the real part of the effective refractive index of the system. In the case of total reflection, ${\cal{R}}(\omega)=1$ and the real part of the electric susceptibility is $-1$. Equating the real part of Eq. (\ref{chi1}) to $-1$ leads to the range of frequencies within which the reflection probability reaches its maximum: $\omega_{01}-\Delta_{1} \leq \omega \leq \omega_{01}+2\Delta_{1}$ and hence the frequency window with the maximum reflection shows a width of  $3\Delta_{1}$ which agrees with the quantum simulation \cite{PhysRevLett.113.163603}. For relatively thin slab of width $\ell \le \lambda_{01}$, the width of the reflection window is noticeably below $3\Delta_{1}$. This is due to the fact that in this analysis we have neglected contributions from multiple reflections that occur inside the slab. For a thin system, contributions from the two different vacuum-medium interfaces are not negligible and obviously affect the reflection spectra.

\begin{center}
    \begin{figure}[h]
     \includegraphics[width=8cm]{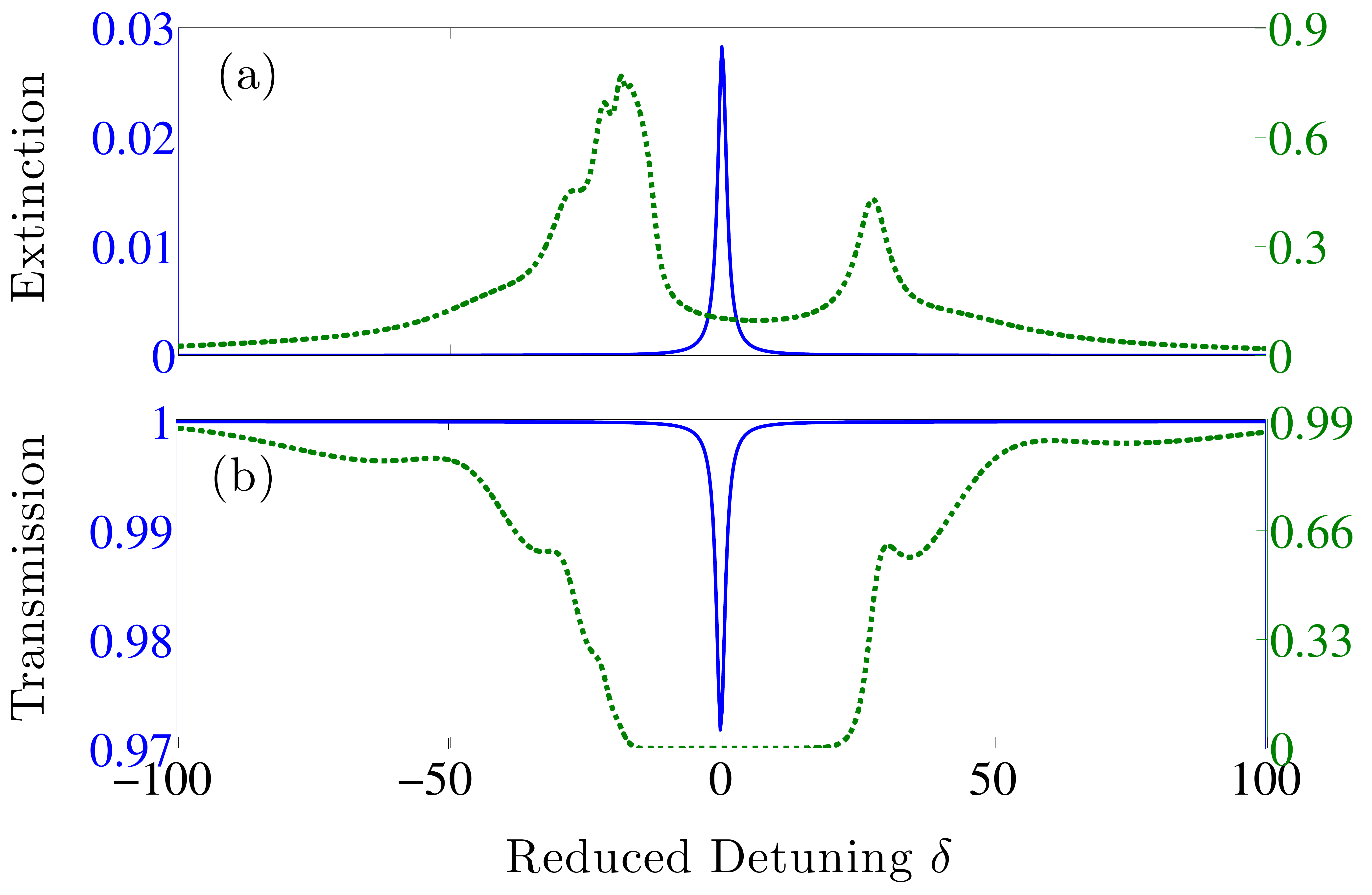}
      \caption{(Color online) Extinction (panel a) and transmission (panel (b)) probabilities as functions of the reduced detuning $\delta$. The solid blue curves (left vertical scale) represent the results for $\Delta_{1}/\gamma = 10^{-3}$  and the dotted green lines (right vertical scale) show the data for $\Delta_{1}/\gamma = 12$. All other parameters are the same as in the Figure \ref{ref_two_level}.}
      \label{Trans_Abs_two_level}
    \end{figure}  
   \end{center}

Fig. \ref{Trans_Abs_two_level}  shows the extinction (panel (a)) and transmission (panel (b)) spectra at $\Delta_{1}/\gamma = 10^{-3}$ (blue solid line) and $\Delta_{1}/\gamma = 12$ (green dashed line). For  $\Delta_{1}/\gamma = 10^{-3}$, almost all the incident energy is transmitted. Clearly in the weak interaction regime, the local polarization barely modifies the local electric field experienced by a given emitter. Hence the emitters respond to the field as if they were independent. As the mutual interaction between the dipoles increases, the system becomes opaque and reflects most of the incident energy  \cite{PhysRevA.61.063815}. In complement with reflection and transmission, the remaining energy is absorbed by the system and decays due to radiationless transitions described by $\hat\Gamma$ in Eq.  (\ref{Opt_Bloch_1}). In particular at $\Delta_{1}/\gamma = 12$, the system loses its transparency, as seen in Fig.  \ref{Trans_Abs_two_level}(b), and behaves almost like a mirror around a broad range of frequencies near the transition frequency.

It should be emphasized that the reflection from strongly interacting dipoles resembles Bragg mirrors (also referred to in the literature as distributed Bragg reflectors)  \cite{ajp1.4832436}. Bragg mirrors are composed of alternating dielectric layers with different refractive indices. The high reflectivity is achieved by multiple reflection from layers of the mirror. This leads to a controllable constructive interference. Spatially varied refractive index profiles determine the fringes in the spectrum. In contrast to Bragg mirrors our system is completely homogeneous. Since we perform simulations in a linear regime, the refractive index of the system is a constant throughout the layer. This rules out the possibility of having multilayer interferences. All the effects observed in the spectra are due to the collective response of the system. The width and the position of the reflection window are characterized by the physical properties of the system such as the transition dipole, transition frequency, material density. The qualitative explanation for the modifications in the spectra when the system enters the strong interaction regime relies on the fact that the polarization of the system is enhanced due to the strong dipole-dipole interactions. Prior to the detailed quantitative analysis of the spectral features, we want to estimate the accuracy of the semi-classical model and compare it with results obtained via exact numerical simulations.
 
\begin{figure}[h]
   \begin{center}
    \includegraphics[width=8cm]{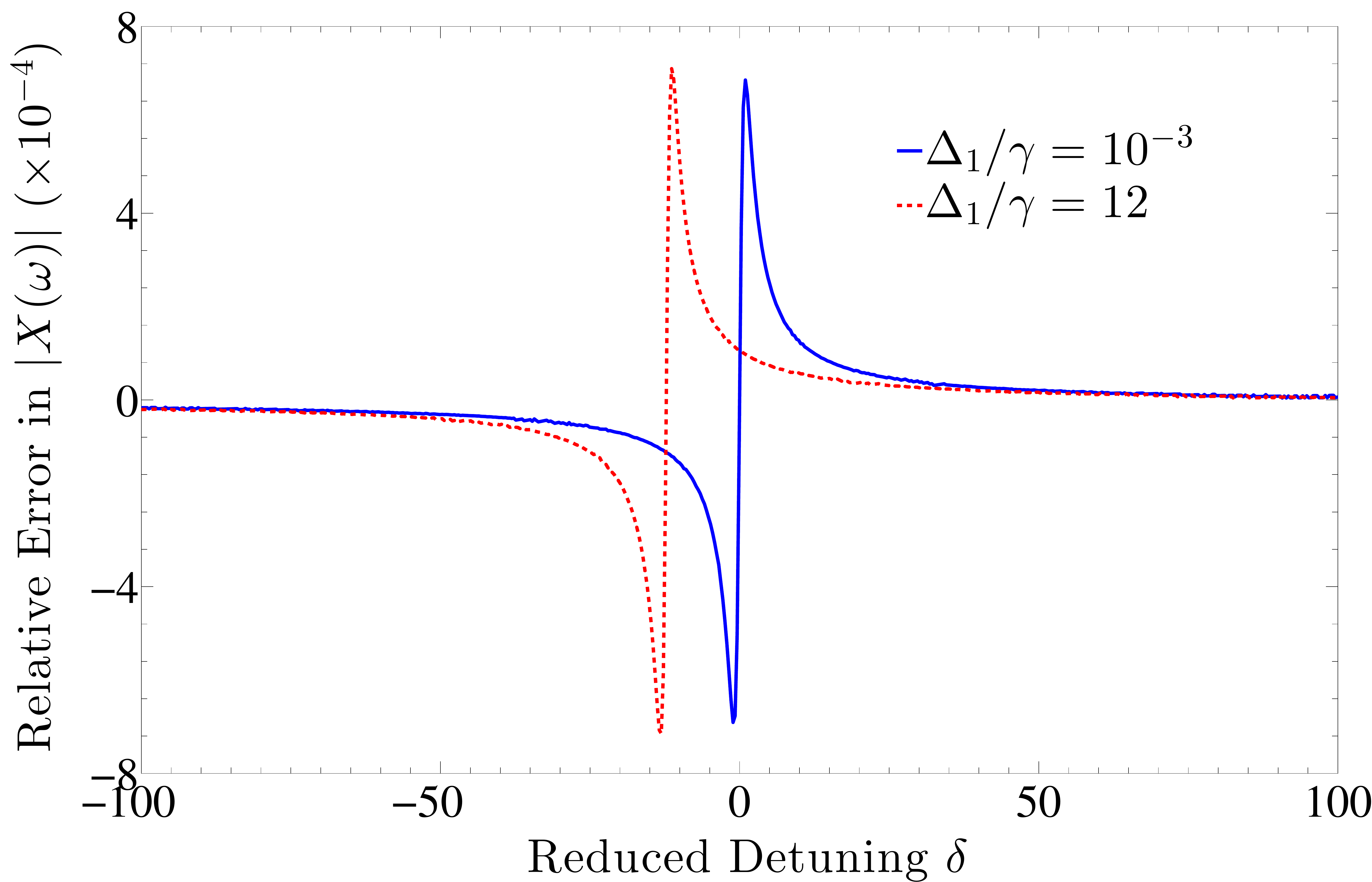}
     \caption{
     (Color online) Relative error in the absolute values of the electric susceptibility using the semi-classical approximation with respect to the numerical integration of the Maxwell-Bloch equations as a function of the reduced detuning  for parameters used in Figure \ref{Trans_Abs_two_level}. The blue solid line shows the relative error for weakly interacting dipoles and the red dashed line presents the same for strongly interacting dipoles. } 
     \label{chi2}
   \end{center}
  \end{figure}

Fig. \ref{chi2} shows the relative errors of the electric susceptibilities for interacting two-level emitters obtained using the semi-classical model with respect to the numerical integration of the Maxwell-Bloch equations, Eq. (\ref{chi1}) in the weak ($\Delta_{1}/\gamma = 10^{-3}$ ) and strong ($\Delta_{1}/\gamma = 12$) interaction regimes. The solid blue line shows the relative error for weakly interacting dipoles and the red dashed line, is the same for strongly interacting dipoles. Both relative errors are peaked around the corresponding transition frequencies with a maximum amplitude of about $0.08\%$ and we conclude that the analytical model is accurate enough to analyze the numerical results.
 
 \subsection{Fabry-P\'erot Modes}

\begin{figure}[h]
   \begin{center}
    \includegraphics[width=8cm]{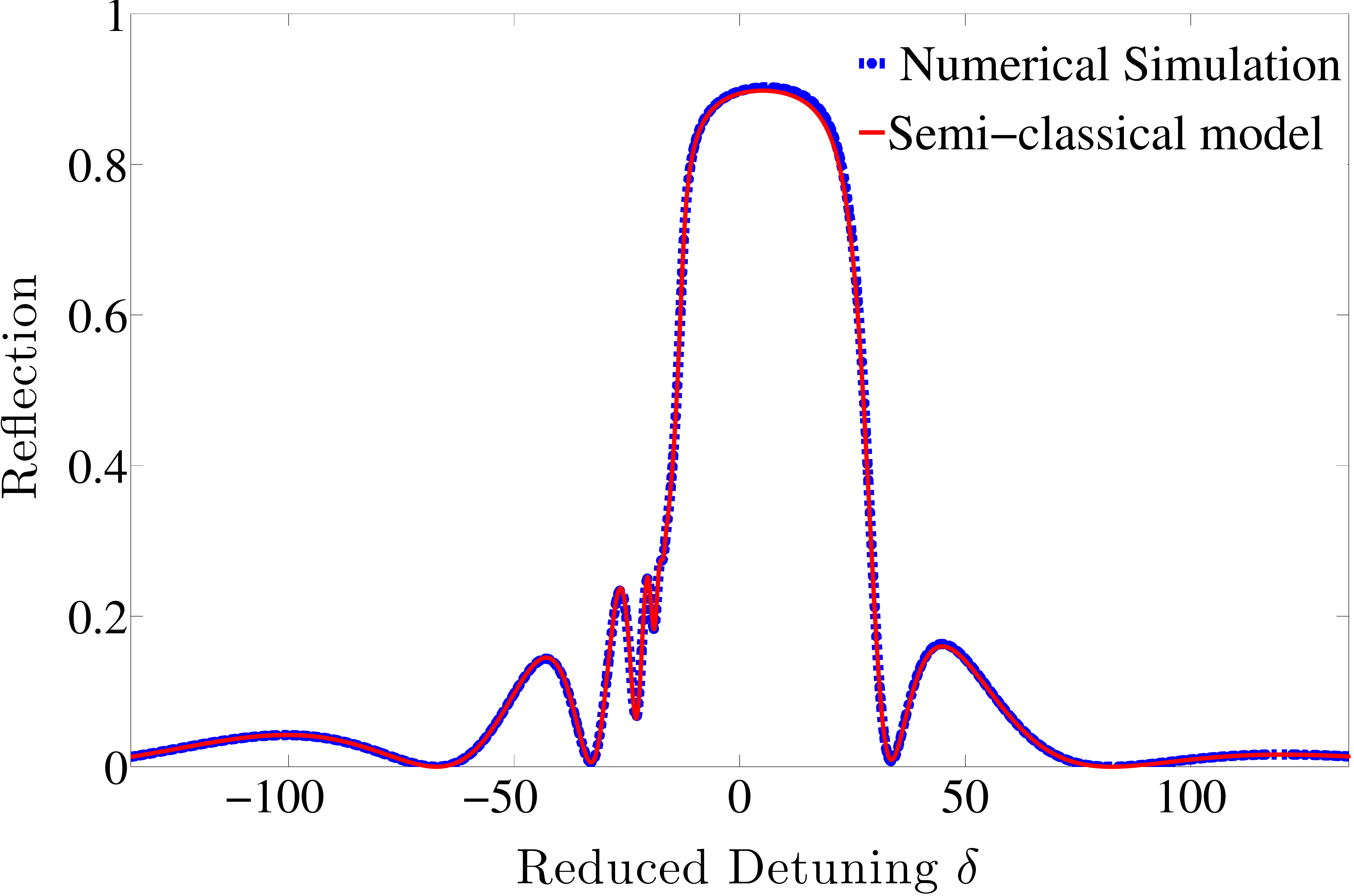}
     \caption{
     (Color online) Reflection as a function of the detuning for $\ell = 600$ nm . The dashed blue line is the reflection probability calculated by integrating the Maxwell-Bloch equations and the solid red line is the  same obtained from the semi-classical approximation (Eq. (\ref{ref_Tran} a)). Other parameters are  $\Delta_{1}/\gamma = 12$, $\mu_1 = 2$ D and $\gamma = 10.5$ THz .} 
    \label{FP_modes}
   \end{center}
  \end{figure}

 The semi-classical model explains the spectral features such as the shift in the resonance frequency and the width of the reflection window  for a system of interacting two-level emitters in the strong interaction regime. The features that are not yet interpreted by the model are sidebands (see Fig.  \ref{ref_two_level} for $\Delta_1/\gamma = 10$). Those however can be interpreted as simple Fabry-P\'erot modes \cite{PhysRevLett.108.173601,fowles2012introduction}. The semi-classical model does not yet contain any information about the geometry of our system, assuming homogeneous distribution of the emitters. With two or more interfaces present, one can expect to observe interferences between reflected and transmitted fields due to the possible multiple reflections inside the slab. In this case, the system can be seen as a Fabry-P\'erot etalon consisting of two parallel partially reflecting interfaces shown as the interfaces 1 and 2 in Fig.  \ref{system}. EM fields at frequencies outside the transmission window are partially reflected and transmitted, resulting in non-zero optical path differences and hence giving rise to an interference pattern as seen in Fig.  \ref{FP_modes} for instance. Introducing the decay of the EM fields inside the slab and taking into account multiple reflections with proper boundary conditions \cite{fowles2012introduction}, we find the reflection $r(\omega)$ and transmission $t(\omega)$ coefficients for the system as
 
  \begin{subequations}
   \label{collective}
     \begin{align}
      & r(\omega) =  \frac{-2i \text{sin}(n(\omega)k\ell) [n(\omega)^2 - 1] e^{-ink\ell}}{[n(\omega)+1]^2 - [n(\omega)-1]^2 e^{-i2n(\omega)k\ell}},  \\
      & t(\omega) =  \frac{4n(\omega) e^{-\kappa \ell}}{[n(\omega)+1]^2 - [n(\omega)-1]^2 e^{-i2n(\omega)k\ell}},  
     \end{align}
   \end{subequations} 
where $n(\omega) = \sqrt{1+X(\omega)}$ is the refractive index, $k= 2\pi/\lambda$ is the  propagation constant of the applied field in the vacuum and $\kappa$ is the imaginary part of the propagation constant inside the medium. The associated reflection and transmission spectra predicted by the semi-classical model are given by

 \begin{subequations}
   \label{ref_Tran}
     \begin{align}
 R(\omega) = |r(\omega)|^2, \\
 T(\omega) = |t(\omega)|^2.
     \end{align}
   \end{subequations} 
  
 The expressions (\ref{ref_Tran}) together with (\ref{collective}) can accurately reproduce the positions of the maxima and minima of the sidebands in the calculated spectra. Fig. \ref{FP_modes} shows the reflection spectrum calculated by integrating Maxwell-Bloch equations (blue dashed line) and the same obtained from the equation (\ref{ref_Tran} a) shown by the solid red line. The two curves are in perfect agreement.
 
\section{Multilevel Systems}
 \label{many-levels}

 Let us consider a ground state coupled to $(N-1)$ excited states. As in the previous case, the infinite slab of emitters is exposed to a transverse electric field. In the strong interaction regime, the Lorentz-Lorenz correction results in redshifts for all transitions. We thus expect to observe $(N-1)$ reflection windows of width $3\Delta_{\jmath}$ each. The most interesting case is when the transitions are significantly overlapping, leading to interference effects. For three level systems, in the presence of such overlapping resonances, we observed DIET which is due to the interference between two indistinguishable excitation pathways \cite{PhysRevLett.113.163603}.
 
\begin{figure}[h]
   \begin{center}
    \includegraphics[width=8cm]{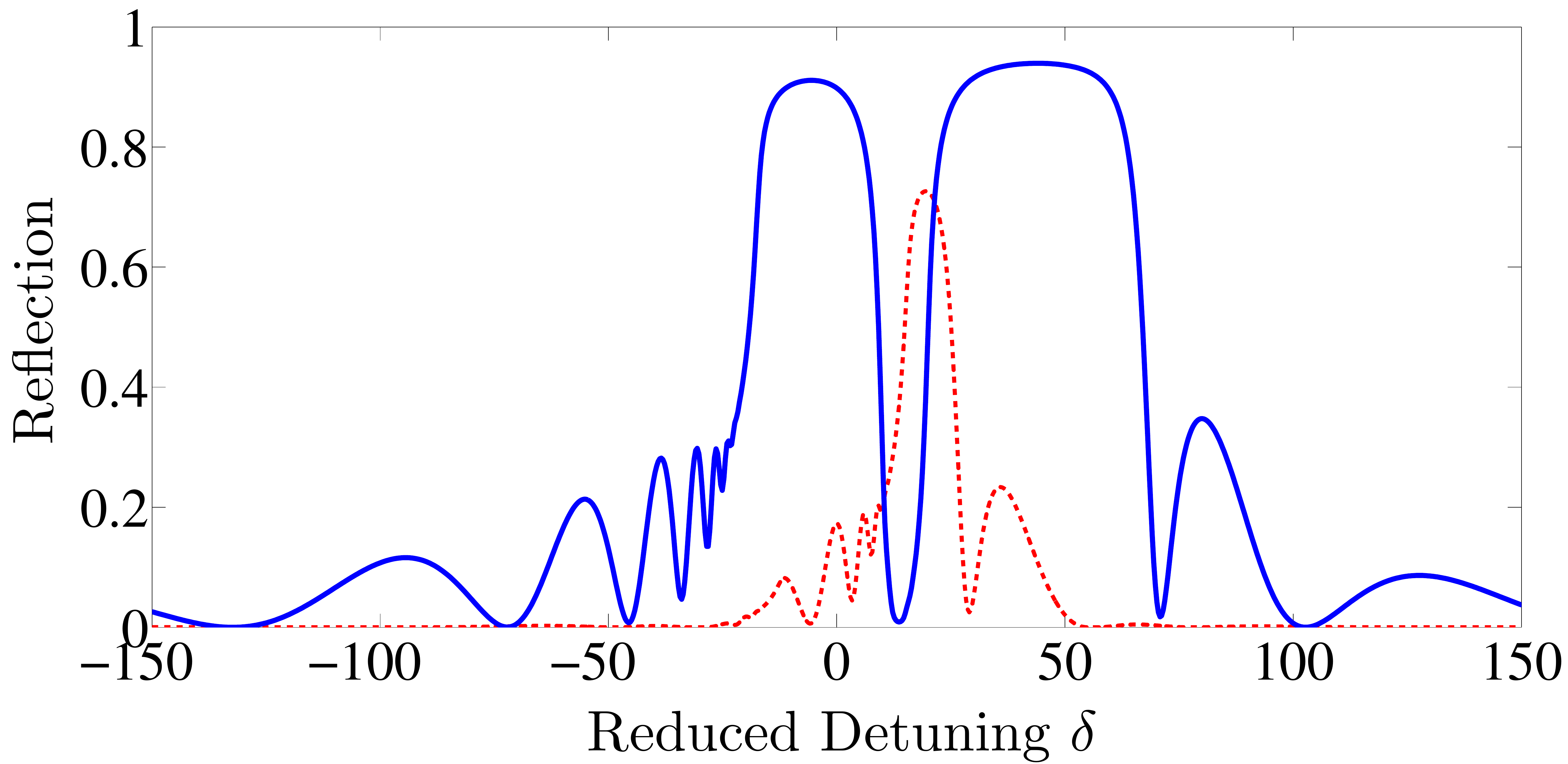}
     \caption{(Color online) The blue solid line is the reflection as a function of the reduced detuning of three-level emitters with $\Delta_{1}/\gamma =\Delta_{2}/\gamma = 12$. The red dotted line is the reflection from the same system in the absence of coupling between the dipoles. The excited states are separated by $\omega_2 - \omega_1 = 4.6\gamma$. The reduced detuning $\delta$, is defined with respect to the transition frequency of the first excited state of this three level system.} 
     \label{Three_level_ref}
   \end{center}
  \end{figure}

  The calculated reflection spectrum for a three level system with overlapping transitions is shown in Fig.  \ref{Three_level_ref} as a blue solid line. It is compared with the reflection spectrum (red dotted line) obtained from a slab of two types of uncoupled two-level emitters having the same physical parameters which is calculated as the product of the reflection probabilities of each dipole. For the uncoupled dipoles the reflection spectrum behaves completely differently compared to the case when they are coupled. For uncoupled dipole systems, the total reflection is peaked where the two independent reflection windows corresponding to each dipole are contributing. This region corresponds to a clear overlap of two resonances.
  
 The strong interaction between the dipoles in the system broadens the transition that results in the overlapping of two closely spaced allowed transitions and hence sets up a competition between two possible excitation pathways. We obtain a narrow window where the excitation probability becomes zero. The presence of a hole in otherwise flat reflection spectra is a clear signature of the destructive interference between the radiation emitted by coupled dipoles and hence a signature of DIET \cite{PhysRevLett.113.163603}. Except for the observed hole in the reflection spectrum, all features discussed in the previous section are clearly present in the three-level system spectra. At the frequency where there is no reflection, we observe a strong transmission if damping processes in the system are not too efficient. The observed transparency, similar to electromagnetically induced transparency (EIT) \cite{ajp1.1412644, PhysRevA.88.053827}, may lead to many potential applications such as slow light discussed later in this paper. The position of the transparency window can be controlled to a certain extend by changing the material parameters \cite{PhysRevLett.113.163603}.
  
  The same type of interference effect can happen if the system has a series of closely spaced energy levels. One of the potential actual systems for studying DIET are Rb atoms. Rb is used experimentally both as a thermal gas  \cite{PhysRevLett.108.173601} and as a cold atomic gas \cite{PhysRevLett.113.133602}. Dipole-dipole interaction depends on the square of the transition dipoles and Rb
atoms are characterized by large $S$ to $P$ transition dipoles \cite{steck}. We consider the D1 transition, \emph{i.e}, $5 ^{2} S_{1/2}$ to  $5 ^{2} P_{1/2}$ of $^{\text{85}}\text{Rb}$. Both  $5 ^{2} S_{1/2}$ and $5 ^{2} P_{1/2}$ are split into two sub-levels due to the hyperfine interaction \cite{HEwhite} and hence there are four dipole-allowed transitions. Thus the D1 transitions of $^{85}\text{Rb}$ isotope can be considered as a superposition of two three-level systems having same excited states but two different ground states. Since the energy splitting between the excited states is smaller than the splitting of the ground states, it gives a possibility to observe DIET between two excited states that are coupled to a specific ground state as well as between two excited states coupled to two different ground states for different atomic densities.
  
  The susceptibility of a system consisting of two ground states  can be written as
  
  \begin{equation}
   \label{chi_for_mixed_dipoles}
    X(\omega) = \frac{{\sum_{\jmath}{\omega^2_{{\cal P} \jmath}/W_{\jmath}}}+{\sum_{\jmath'}{\omega^2_{{\cal P} \jmath'}/W_{\jmath'}}}}{{1-\frac{1}{3}\big{(}{\sum_{\jmath}{\omega^2_{{\cal P} \jmath}/W_{\jmath}}}+{\sum_{\jmath'}{\omega^2_{{\cal P} \jmath'}/W_{\jmath'}}}}\big{)}},
  \end{equation}
  where the index $\jmath$ corresponds to the transitions from the sub-level $F=2$  and  $\jmath'$ is for the transitions from $F=3$. It is important to note that Eq.  (\ref{chi_for_mixed_dipoles}) can also be used to analyze systems composed of different types of atoms or molecules. The LL shifts for D1 transitions of Rb can be written as
  
  \begin{equation}
   \label{LL_shift_Rb}
    \Delta_{FF'} = \Delta_{0} S_{FF'}^2,
  \end{equation}
  where $\Delta_{0} = n_0\mu_{0}^2/9\hbar\varepsilon_0$ is the LL shift defined in the absence of hyperfine splitting and $S_{FF'}$ is a measure of relative strength of the transitions \cite{steck}.
  
\begin{figure}[h]
   \begin{center}
    \includegraphics[width=8.5cm]{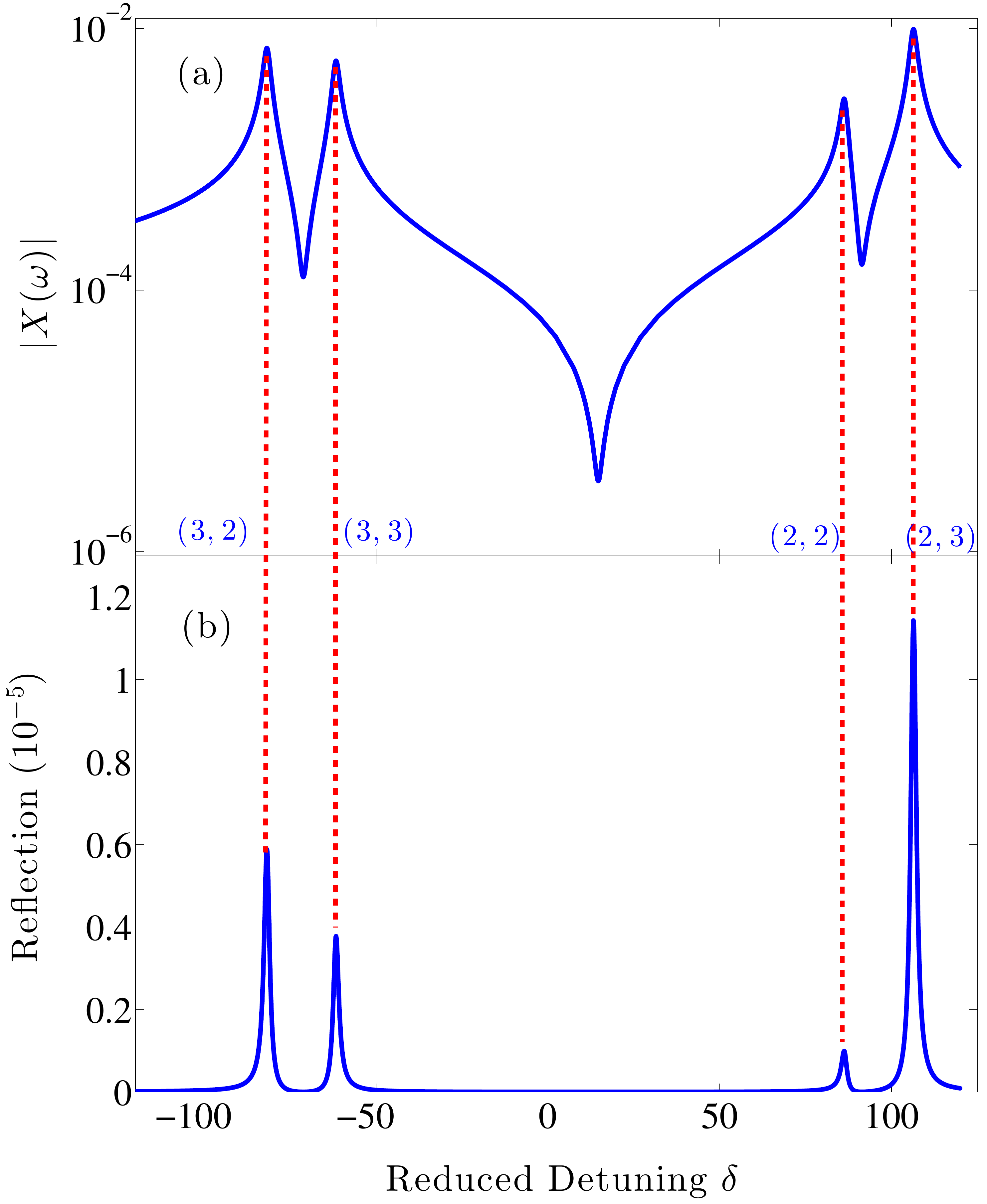}
     \caption{(Color online) Susceptibility (panel (a), log scale) and reflection probability (panel (b), linear scale) as a function of the reduced detuning with $\Delta_{0}/\gamma = 3.8\times 10^{-3}$ . The reduced detuning $\delta$ is defined with respect to the transition frequency of $5 ^{2} S_{1/2}$ to  $5 ^{2} P_{1/2}$ in the absence of hyperfine splitting ($377.107$ THz) and $\gamma = 36.13$ MHz. The positions of transitions $F= 2,3\rightarrow F' = 2,3$ are indicated by the red dashed lines marked with $(F,F')$.} 
     \label{Rb-85-weak}
   \end{center}
  \end{figure}

  The reflection and susceptibility of a slab of thickness $\ell = 600 $ nm comprised of $^{85}\text{Rb}$ for $\Delta_{0}/\gamma = 3.8\times 10^{-3}$ are shown in Fig.  \ref{Rb-85-weak}. The reduced detuning $\delta$ is defined with respect to $377.107$ THz, the D1 transition frequency in the absence of hyperfine splitting \cite{steck}.

   The double peaks in the negative detuning region are due to the transitions $F=3 \rightarrow F'=2$ and  $F=3 \rightarrow F'=3$ and those in the positive detuning region are due to the transitions $F=2 \rightarrow F'=2$ and  $F=2 \rightarrow F'=3$. The separation of the double peaks is due to the hyperfine splitting of the ground state. Since the system is in the weak interaction regime, the transitions are independent and they are not overlapping. It explains the reflection spectrum in  Fig. \ref{Rb-85-weak}(b) with a series of Lorentzian profiles corresponding to different transitions. 
   
\begin{figure}[h]
   \begin{center}
    \includegraphics[width=8.5cm]{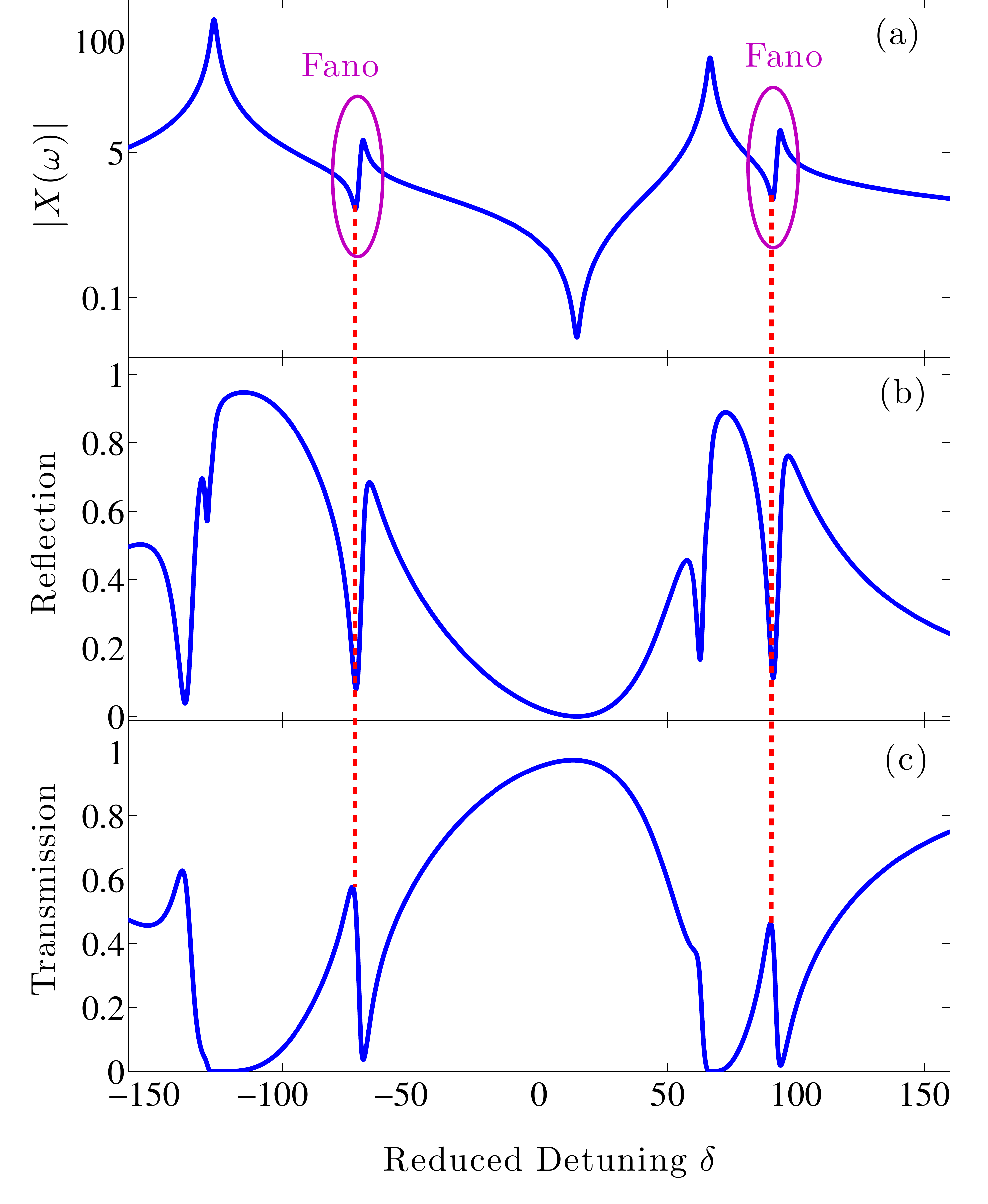}
     \caption{(Color online) Susceptibility (panel (a), log scale), reflection (panel (b), linear scale) and transmission (panel (c), linear scale) as functions of the reduced detuning of $^{85}\text{Rb}$ with $\Delta_{0}/\gamma = 21$ .  Vertical red dashed lines show the positions of destructive interference between the different transition dipoles.} 
     \label{Rb-85-strong1}
   \end{center}
  \end{figure}

Since the dipoles are relatively large for D1 transitions, relatively small increments in the gas pressure will couple the different transitions and hence lead to the DIET regime. The calculated reflection and transmission spectra and the susceptibility for such a system with $\Delta_{0}/\gamma = 21$ are given in Fig. \ref{Rb-85-strong1}. Panel (a) shows the susceptibility, panel (b) presents the reflection spectrum, and panel (c) shows the transmission.
   
  Overlapping of the transitions to $F'=2$ and $F'=3$ from the ground states result in two Fano profiles in the susceptibility. Two sharp minima in the reflection spectrum are the signatures of destructive interferences between the transition dipoles. Concurrently we see two narrow peaks in the transmission spectrum. These frequencies at which DIET takes place are shown as red dashed lines in the Figure.

\begin{figure}[h]
   \begin{center}
    \includegraphics[width=8.5cm]{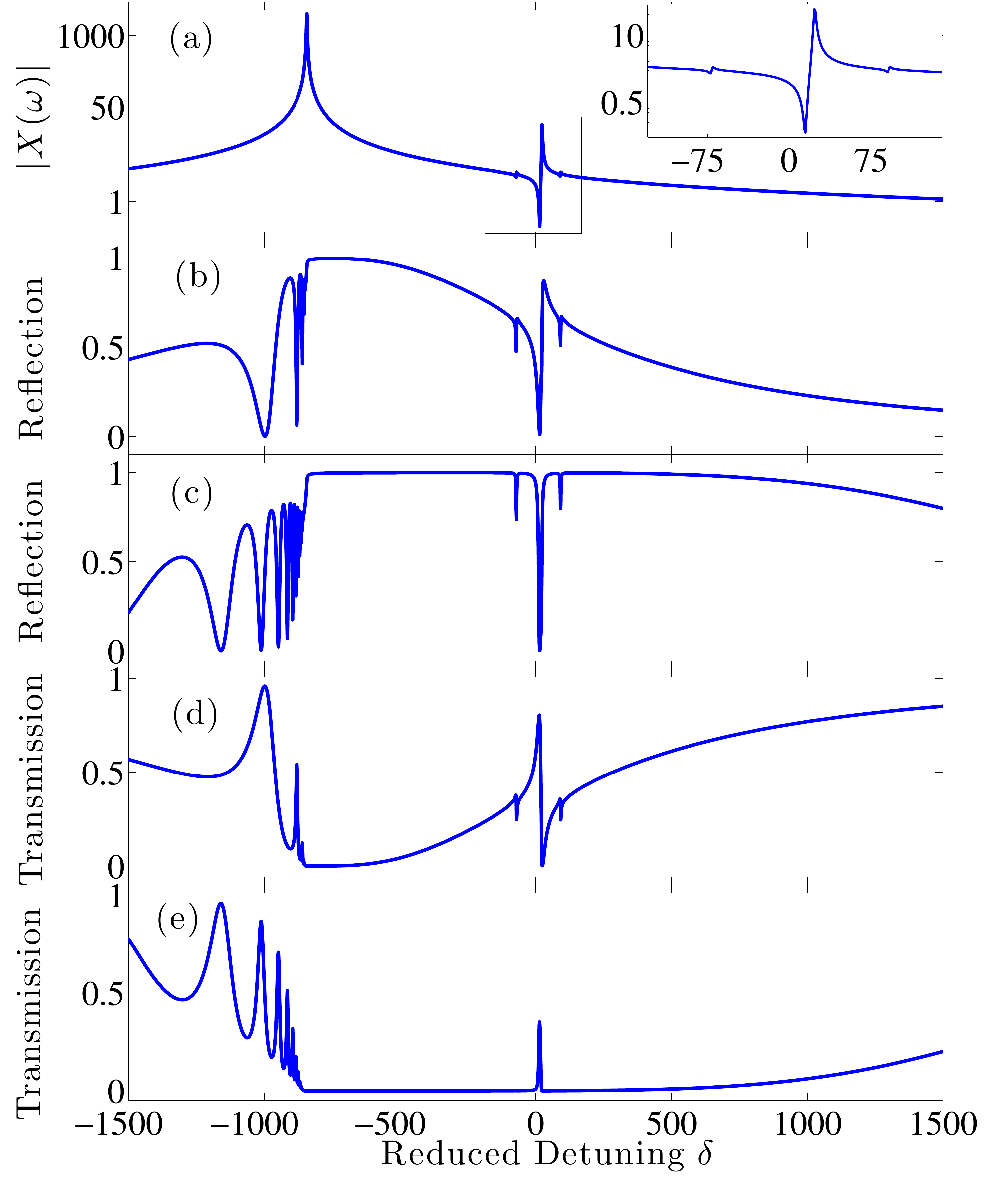}
     \caption{(Color online) Susceptibility (panel (a), log scale) of a slab of $^{85}\text{Rb}$ atoms with $\Delta_{0}/\gamma = 200$. Reflection spectra  (panel (b) and c) for thickness $\ell=600$ nm and $\ell=2.5$ $\mu\text{m}$  and corresponding transmission spectra in panel (d) and (e) as functions of the reduced detuning. A zoom of the susceptibility is given in the inset of panel (a) for better view of the faint Fano profiles associated with the holes in the reflection spectra.} 
     \label{Rb-85-strong2}
   \end{center}
  \end{figure}

  Further increase of the dipole-dipole interaction mixes the  $F'=2$ and $F'=3$ excited states into a single excitation state which is coupled to two ground states ($F=2$ and $F=3$ of $5^2S_{1/2}$). It leads to DIET due to the overlapping of transitions to the excited state from two different ground states. Fig. \ref{Rb-85-strong2} shows such an effect in strong interacting samples of $^{85}\text{Rb}$ with $\Delta_{0}/\gamma = 200$. Panel (a) in Fig.  \ref{Rb-85-strong2} is the modulus of the susceptibility, panel (b) and (c) are the reflection spectra for slabs of thicknesses $\ell = 600$ nm and $\ell = 2.5$ $\mu$m, respectively, and panels (d) and (e) are the corresponding transmissions. 
  
  The susceptibility shows one dominant and two faint Fano type profiles. The latter clearly shown in the inset in panel (a) are due to the splitting of the excited states. The dominant Fano profile corresponds to the overlapping of two coupled dipoles $F = 3 \rightarrow F'= 2,3$ and   $F = 2 \rightarrow F'= 2,3$. It leads to DIET which is well resolved in the reflection (Fig. \ref{Rb-85-strong2}(b) and (c)) and transmission (Fig. \ref{Rb-85-strong2}(d) and (e)) spectra of slabs of thickness $600$ nm and $2.5$ $\mu$m. This can be understood if we assume that the D1 transitions are a mixture of two three-level systems having the same excited states. In such a picture, the dominant DIET in the strong interaction regime can be interpreted as an interference effect between the transitions to a single excited state from two different ground states. The two small Fano profiles in the susceptibility are due to the splitting of the excited states. They also result in DIET which explains two small transmission peaks on the left and right of the dominant DIET signal from the $600$ nm slab. These two transmission peaks are not resolved for the $2.5$ $\mu$m slab due to damping effects. The system, therefore, acts almost as if it was comprised of three-level emitters \cite{PhysRevLett.113.163603}. However the splitting of the excited state leaves spectral holes in the reflection spectra (Fig. \ref{Rb-85-strong2}(c)).

  \section{Potential Applications}
\label{applications}
\begin{figure}
   \includegraphics[width=8.5cm]{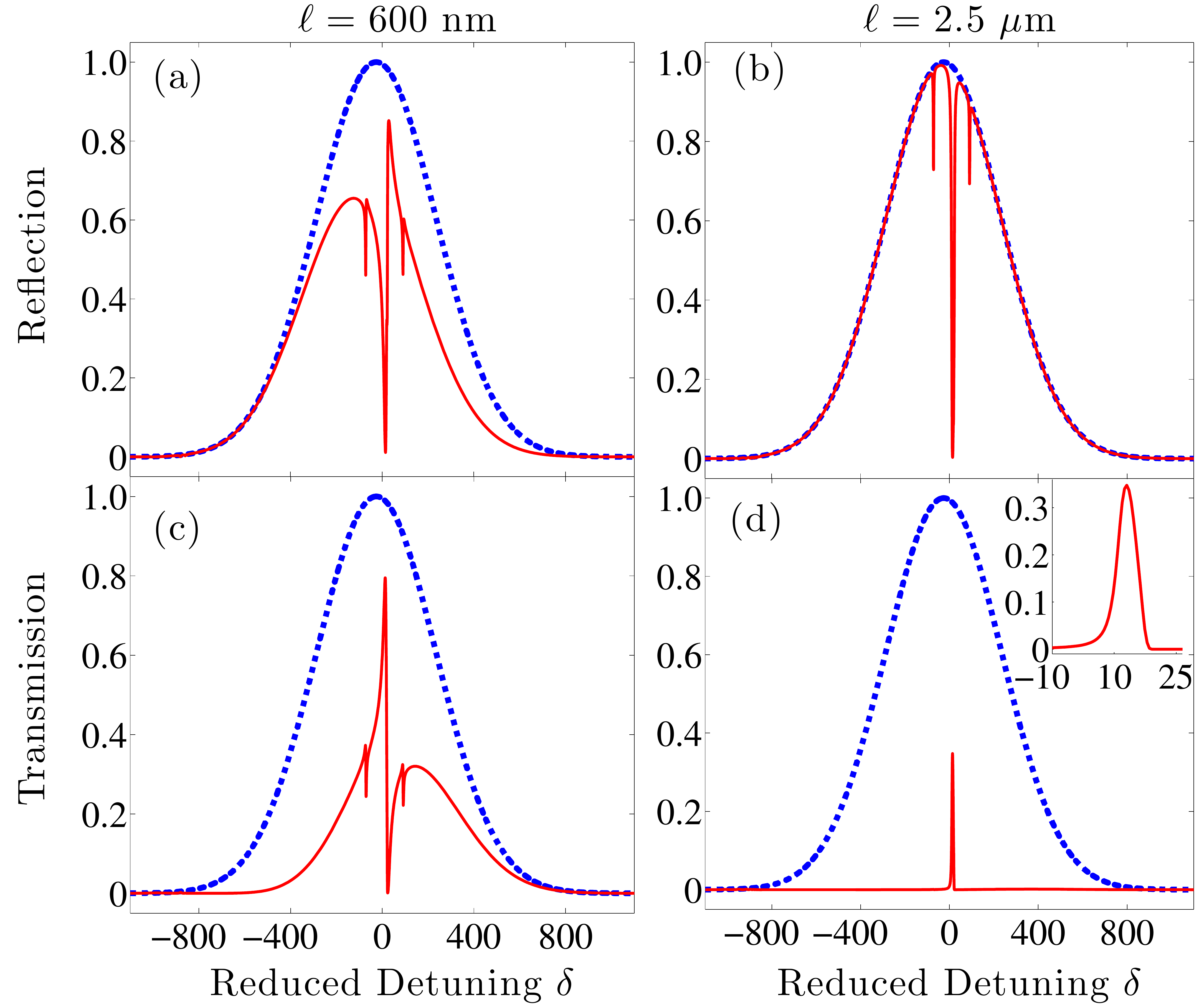}
    \caption{(Color online)Reflected (panels (a) and (b)) and  transmitted (panels (c) and (d)) pulse shapes from the layer of $^{85}\text{Rb}$ atoms  at $\Delta_{0}/\gamma = 200$ as functions of the reduced detuning $\delta$. The blue dashed lines in the panels are the normalized incident pulse and the red lines are the reflected and transmitted pulse envelopes. Panels (a) and (c) are for a slab of $\ell = 600$ nm  and panels (b) and (d) are for $\ell=2.5$ $\mu$m. }
   \label{pulse_shape}
  \end{figure}

The destructive interference of two overlapping transitions can be used to modify the spectral envelope of an incident pulse, \emph{i.e.} as a pulse shaper. If we consider a slab of interacting multi-level quantum emitters with a reflection window wider than the FWHM, $\delta \omega$ of the incident pulse, the collective response of the quantum emitters induce a selective reflection and transmission of the incident pulse. This is illustrated in Fig. \ref{pulse_shape}  for the D1 transitions of $^{85}\text{Rb}$ at $\Delta_{0}/\gamma = 200$, with a pulse of FWHM, $\delta \omega = 4\pi$ GHz. Panel (a) in Fig. \ref{pulse_shape} compares the normalized reflected pulse envelope (red line) with the incident pulse shape (blue dashed line) of such a pulse with the central wavelength $794.98$ nm for a $600$ nm thick slab of $^{85}\text{Rb}$. Panel (b) is the same as panel (a) but for a $2.5$ $\mu$m thick slab. The normalized transmitted pulse envelopes (red line) are shown in panel (c) for $600$ nm and in panel (d) for $2.5$ $\mu$m. A zoom around the central wavelength is given in the inset of panel (d) to see the shape of the pulse transmitted though the 2.5 $\mu$m slab.

 From the Figs. \ref{pulse_shape}(a) and (b) it is clear that the reflection from the $600$ nm slab is not as efficient as the reflection from the $2.5$ $\mu$m slab since this slab transmits a large part of the incident pulse (Fig. \ref{pulse_shape}(c)). It is due to the mismatching between the slab thickness $\ell=600$ nm and the transition wavelength $\lambda_0 \simeq 795$ nm that creates long tails of Fabry-P\'erot modes near the reflection window which modify the transmission probability (see Fig. \ref{Rb-85-strong2}(b) and (d)). But for the $2.5$ $\mu$m slab, the condition $\ell > \lambda_0$ is satisfied leading to the wide reflection window. This minimizes the transmission of the pulse of width $\delta \omega < 3\Delta_0$ through the slab except at the positions of DIET. 
 
If the system is comprised of emitters with many optically active excited states that can be populated by the incident laser pulse (such as molecules with ro-vibrational levels for instance \cite{jcp1.4774056}), each transition interacts with the others and interferes constructively or destructively at different frequencies leading to reflected pulses with more than a single frequency amplified or removed from the spectrum. In particular, spectra of the the transmitted pulse in Fig. \ref{pulse_shape}(c) and (d) show that DIET can vary dramatically depending on the system parameters.
 
 \begin{center}
 \begin{figure}[h]
  \includegraphics[width=7.5cm]{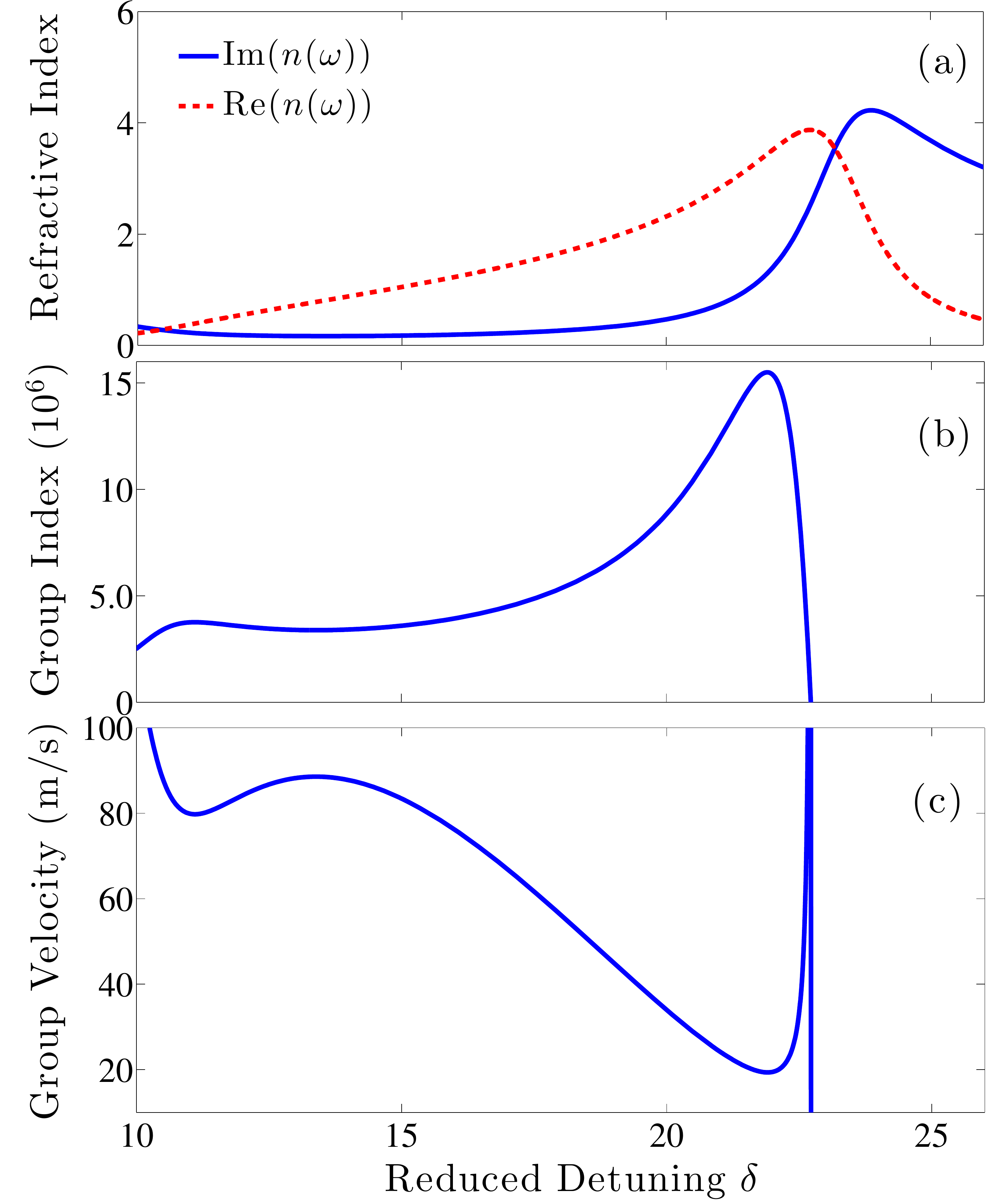}
   \caption{(Color online) Slow light effect associated with  DIET in D1 transitions of $^{85}\text{Rb}$ for  $\Delta_{0}/\gamma = 200$. Panel (a) shows the refractive index: the solid blue line is the imaginary part and the dashed red line is the real part, panel (b) shows the group index of the layer and panel (c) shows the group velocity as functions of the reduced detuning $\delta$.}
   \label{slow_light}
 \end{figure} 
\end{center}

In systems comprised of interacting multilevel emitters we observe a minimum in the reflection window with a highly nonlinear dispersion (see Fig. \ref{Rb-85-strong1} and \ref{Rb-85-strong2}).  It opens several intriguing applications, including slow light \cite{nature09933,PhysRevLett.112.167401}. Fig.  \ref{slow_light} shows the slowing down of the transmitted pulse in a layer of $^{85}$Rb with $\Delta_{0}/\gamma = 200$. Panel (a) shows the refractive index. The blue solid line is the imaginary part and the red dashed line is the real part of the refractive index. Panel (b) shows the group index $n_{g}(\omega) = \text{Re}[n(\omega) + \omega(dn/d\omega)]$ and  the group velocity $v_{g}(\omega) = c/n_{g}(\omega)$ is shown in panel (c).

The system has a very large group index $n_g$ within the window of frequency where DIET occurs (see the Fig.  \ref{slow_light}(d) and the inset of Fig. \ref{pulse_shape}(b)). It peaks at $1.55\times 10^{7}$ similar to a recent experiment on slow light \cite{PhysRevLett.112.167401} and the group velocity (panel (c)) drops below $20$ m/s. But in the system we consider this lowest velocity may not be observed since the imaginary part of the refractive index (see panel (a) in Fig. \ref{slow_light}) is large that set the system as a lossy medium of light at this frequency. It also explains the quick drop in the transmission pulse shown in the inset of Fig. \ref{pulse_shape}(d). The transmitted pulse peaked around the reduced detuning $\delta = 13$ dies out at $\delta = 20$. Within this short window of the reduced detuning the group index $n_g$ is of the order of $10^{6}$ and the velocity is decreased down to $40$ m/s.
 Within this window, the imaginary part of the refractive index is small enough to allow the experimental observation of slowed down transmitted pulse.   
 \section{Conclusion}
\label{conclusion}

In this manuscript, we study collective effects in systems of interacting multilevel quantum emitters and their dependence on different physical parameters such as density. Such effects are studied both numerically using Maxwell-Bloch equations and analytically using a semi-classical model.

When damping is sufficiently low, as in atomic or molecular vapors at low temperatures, for instance, narrow transmission windows are observed. Such narrow windows in an otherwise completely opaque material are due to quantum interferences between different dipoles corresponding to different induced transitions. A clear evidence of such an interaction is the presence of Fano profiles in the susceptibility of the system. It is shown that this effect is amplified when the the dipole-dipole coupling increases. We call the transparency of the system induced by the coupling between different kinds of dipoles \textit{Dipole Induced Electromagnetic Transparency} (DIET). We note that DIET is similar to the well-know electromagnetic induced transparency (EIT)   \cite{ajp1.1412644,PhysRevA.88.053827} the former however is fundamentally different from EIT. EIT requires a strong laser pulse that couples two different quantum states, while DIET is inherently internal, \emph{i.e.} requires overlapping resonances induced by strong dipole-dipole couplings.

The presence of additional levels such as ro-vibrational levels in molecules for instance or multiple optically active electronic states in case of atoms can change the response of the system due to the presence of additional energetically allowed transition dipoles which leave their signatures as spectral holes in the reflection spectrum and as narrow transmission peaks in the transmission spectrum. Such windows are induced via coherent cancellation of the dipoles through a destructive quantum interference effect. In addition, we have shown that this effect can be controlled by changing the number density of quantum emitters. This destructive interference of two or more transition dipoles can be used for shaping laser pulses or for slow light.

\section*{Acknowledgement}
  The authors acknowledge support from the EU (Grant No. ITN-2010-264951 CORINF) and from the Air Force Office of Scientific Research (Summer Faculty Fellowship 2013)
  
 
\bibliographystyle{apsrev4-1}
 
\end{document}